# Monte Carlo Simulation and Experimental Characterization of a Dual Head Gamma Camera

S. Rodrigues[1,2], B. Tomé[3], M. C. Abreu[1,2], N. Santos[4], P. Rato Mendes[5], L.Peralta[3,6]

1. LIP-Algarve, 8005-139 Faro, Portugal
2. University of Algarve, 8005-139 Faro, Portugal
3. LIP, 1000-149 Lisboa, Portugal
4. Siemens Medical Solutions, 2720-093 Amadora, Portugal
5. CIEMAT, 28040 Madrid, Spain
6. FCUL, 1749-016 Lisboa, Portugal
Email: srodrigues@ualg.pt

ABSTRACT

The GEANT4 Monte Carlo simulation and experimental characterization of the Siemens E.Cam Dual Head gamma camera hosted in the Particular Hospital of Algarve have been done. Imaging tests of thyroid and other phantoms have been made "in situ" and compared with the results obtained with the Monte Carlo simulation.

KEYWORDS: Gamma camera, SPECT

## 1. Introduction

Nuclear medicine is a field in which the use of Monte Carlo simulation techniques is rapidly growing, particularly in quality control and dosimetry. The statistical nature of the mechanisms of photon emission, interaction and detection requires the use of Monte Carlo statistical methods in order to achieve the desired levels of detail and accuracy. The current availability of considerable computing power at a low cost also contributes to the generalized growing use of these methods.

One of the most frequent applications in nuclear medicine imaging is detector modeling with three following main purposes: study the interactions within the radiation sensor for each photon and thus correct for sources of image degradation; evaluate techniques of image treatment to quantify the effects of dispersion and attenuation, and thus optimize reconstruction algorithms; and patient dosimetry calculations.

In this work we used the Geant4 simulation toolkit for particle physics and other applications [1]-[3]. It has been first applied in high-energy physics studies at CERN (European Laboratory for Particle Physics) and is being extensively used in medical physics, namely in radiation protection, radiology, radiotherapy, dosimetry, brachytherapy and nuclear medicine imaging [4] [5]. The Geant4 program is very flexible and has the capability to simulate a wide variety of physical processes and three-dimensional complex geometries.

In the following we present the simulation of the Siemens E.Cam Dual Head gamma camera [6] installed in the Nuclear Medicine service of Unidade de Intervenção Cardiovascular (Cardiovascular Intervention Unit), located at Hospital Particular do Algarve (Algarve Private Hospital) in Alvor, Portugal. This gamma camera is used for planar and SPECT clinical imaging, and the aim of the present work is to use Monte Carlo simulation to check on some of gamma camera characteristics announced by the manufacturer.

## 2. Description of the Gamma Camera

The Siemens E.Cam Dual Head gamma camera features two detector heads (figure 1), each consisting of a removable Low Energy High Resolution (LEHR) collimator, a NaI(Tl) (sodium iodide doped with thallium) scintillation crystal, a light guide and an array of photomultiplier tubes (PMTs) as shown in figure 2. The LEHR collimator features parallel holes (2.45 cm height, 0.16 mm septal thickness) with hexagonal cells of 1.11 mm diameter and is used with low energy sources such as $^{99m}$Tc. The NaI(Tl) scintillating crystal is a 9.5 mm thickness single planar crystal with a light yield of about 40000 photons per MeV of deposited energy and an emission spectrum peaked at 415 nm [7].

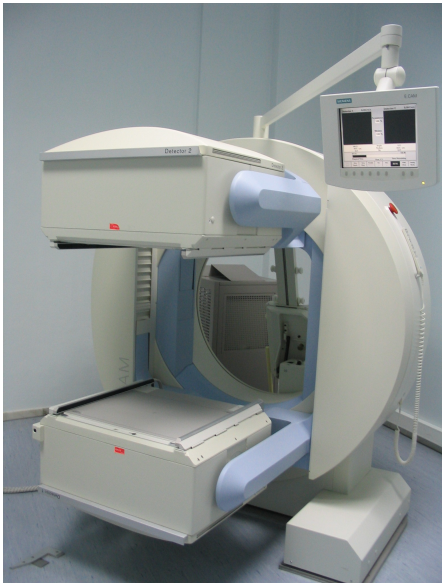

**Figure 1.** Photo of the Siemens E.Cam Dual Head gamma camera.

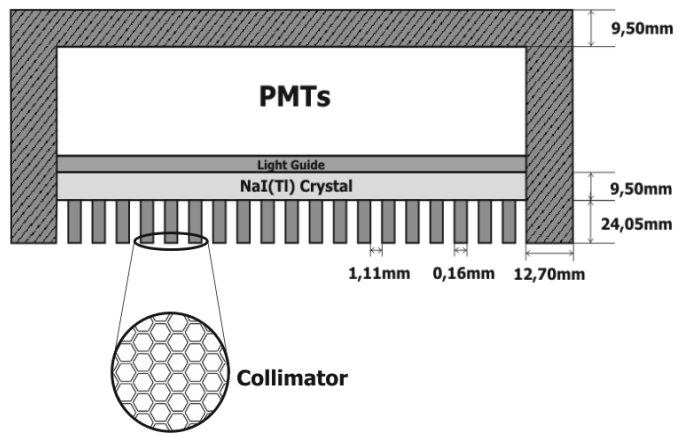

**Figure 2.** Schematic drawing of one detection head of the Siemens E.Cam Dual Head gamma camera.

The light generated in the crystal is collected by a matrix of 59 PMTs, of which 53 are 7.6 cm and 6 are 5.1 cm in diameter. The photocathode is a bialkali type with a quantum efficiency of approximately 30% for the wavelength of maximum NaI(Tl) emission [8]. A light guide ensures a good optical coupling between the scintillating crystal and the PMTs.

## 3. Implementation of the Simulation in Geant4

All geometric characteristics of the gamma camera, radioactive source parameters, particles and processes involved, together with the storage of information for subsequent data analysis were implemented in the Geant4 code in order for the simulations to be as realistic as possible [9]. The physical processes involving primary photon and secondary electron interactions (photoelectric effect, Compton and Rayleigh scattering, Bremsstrahlung and ionization) were defined using the electromagnetic low energy package of Geant4 [10]. The scintillation process was used for generation of optical photons within the NaI(Tl) crystal. An isotropic source emitting 140 keV photons was used for the radioisotope $^{99m}$Tc.

## 4. Performance Evaluation of the Gamma Camera

In this work we present results on the following properties of the Siemens E.Cam Dual Head gamma camera: energy resolution, spatial resolution and linearity. Except for energy resolution other situations correspond events selected within an energy window centered on the $^{99m}$Tc photopeak (130 – 149 keV). The images were reconstructed in matrices of 1024 × 1024 pixels, with a pixel size of 0.6 mm.

### 4.1 Energy resolution

The energy resolution of the gamma camera was measured with a $^{99m}$Tc point source (activity of 1.8×10$^6$ Bq) placed at the center of the field of view (FOV), 10 cm apart the crystal surface. The energy spectrum was acquired for 2 minutes. This setup was simulated for photons generated with a normal incidence to the crystal in order to simulate an ideal collimator. The simulated and the experimental energy spectra obtained are shown in figure 3 and figure 4, respectively. Some differences may be observed between them, the most striking being that the experimental spectrum presents a wider peak which may be explained by the superposition of the energy peaks corresponding to the x-ray escape of $^{53}$I present in the NaI(Tl) crystal (~110 keV), the $^{99m}$Tc photopeak (140 keV) and the sum of 140 keV with the x-ray energy of $^{99m}$Tc (total ~160 keV) [11], which cannot be separated by the detector.

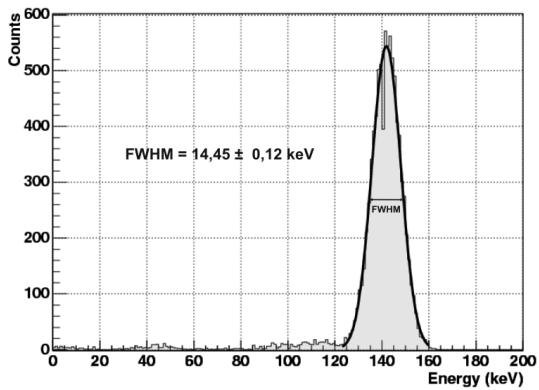 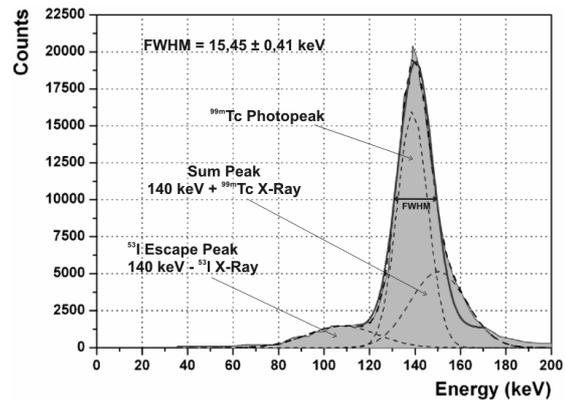

**Figure 3.** Simulated energy spectrum of a monoenergetic gamma point source of 140 keV with normal incidence on crystal without collimator.

**Figure 4.** Experimental energy spectrum of $^{99m}$Tc, decomposed on three Gaussian functions corresponding to different sum and escape energy peaks.

The results of fitting the peaks in both spectra by Gaussian functions gave a FWHM of 14.45±0.12 keV for the simulation and 15.45±0.41 keV for experimental data, corresponding to an energy resolution of 10.3±0.1 % for the simulation and 11.0±0.3 % for the experimental measurements. The manufacturer provides an upper limit of 9.9 % FWHM for the energy resolution at 140 keV [12].

4.2 Sensitivity

The sensitivity of the gamma camera is determined by taking the ratio of the detected photons in the selection window to the total number of photons emitted in the solid angle of the FOV. The sensitivity, was experimentally measured and calculated by simulation for a $^{99m}$Tc point source centered on FOV and located at the following distances from the detector with the LEHR collimator attached: 0, 10, 15 and 20 cm. The experimental measurement was performed with a source activity of 1.9×10$^6$ Bq during 2 minutes whereas for the simulation 33 million photons were generated for each source position. The radioactive decay time was taken into account in experimental measurements and background radiation was subtracted. The results obtained for sensitivity for both simulations and real data are plotted in figure 5 for four source-collimator distances. The systematic higher sensitivity value obtained by the simulation is due to the acquisition dead time and signal overlap in the detector, which was not accounted for in the Monte Carlo.

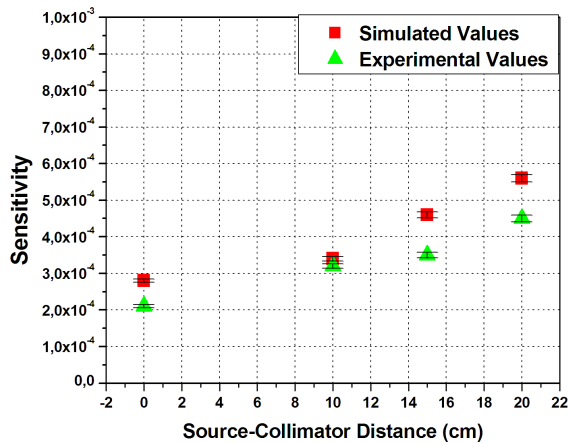 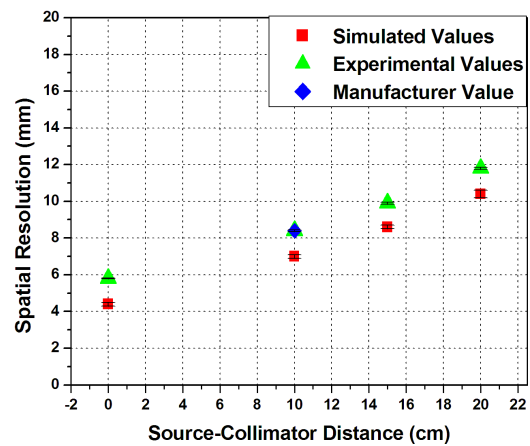

**Figure 5.** Comparison between simulated and experimental sensitivities for a $^{99m}$Tc point source at four source-collimator distances.

**Figure 6.** Comparison between simulated and experimental spatial resolution for a $^{99m}$Tc point source located at four different distances from the collimator. The manufacturer value at 10 cm is also presented.

4.3 Spatial Resolution

The experimental measurements were done using a 1.9×10$^6$ Bq $^{99m}$Tc source with a diameter of 2 mm, placed at the center of FOV. Acquisitions for distances of 0, 10, 15 and 20 cm from source to the face of the LEHR collimator, 2 minutes each were obtained. The spatial resolution of the gamma camera was also obtained by Monte Carlo simulation using point spread functions (PSF) at the same distances as the experimental data.

The FWHM values of point spread functions of simulation and experimental measurements are presented in figure 6. These values were obtained for four distances between the source and the collimator surface. We observed a difference between simulated and experimental results that is constant over the range of the source-collimator distance selected. This difference is due to the fact that the point source used in simulation is negligible extent compared to the 2 mm diameter source used in experimental measurements. For a source-collimator distance of 10 cm, the simulated and experimental spatial resolution are 7.0±0.1 mm and 8.4±0.1 mm, respectively. The value provided by manufacturer for a point source at this distance is 8.3 mm [12], which nicely agrees with our experimental value.

4.4 Imaging Evaluation

The basic process to reconstruct the incident photon position in the crystal is to compute the position centroid ($x_c, y_c$) using the PMTs coordinates ($x_i, y_i$) and the collected signals $w_i$:

$$x_c = \frac{\sum w_i x_i}{\sum w_i} \quad \text{and} \quad y_c = \frac{\sum w_i y_i}{\sum w_i} \, . \tag{1}$$

It is well known that this simple procedure has several bias, that for instance privileges the PMTs coordinates, accumulating the reconstructed points around these positions. In this work a simple procedure to partially correct this effect was tested. In figure 7 it is represented the reconstruct $y$ position as function of the Monte Carlo simulated $y$ coordinate of the incident photon along the straight line x=0, on the surface of the collimator. This plot can be transformed in a simpler periodic function if instead we display the difference $y_c$-$y$ as function of $y$ (figure 8). Now straight lines easily approximate the function. These fits can be used to correct the reconstructed position, leading to a much better linear correlation between the $y_c$ and $y$ as shown in figure 9.

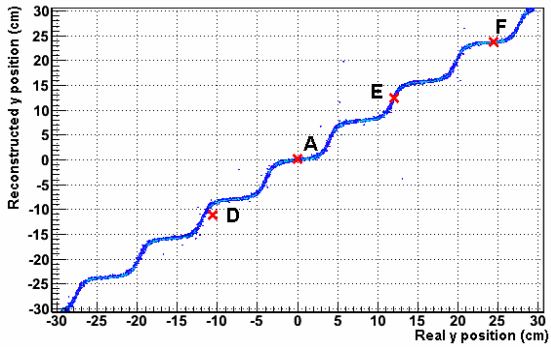 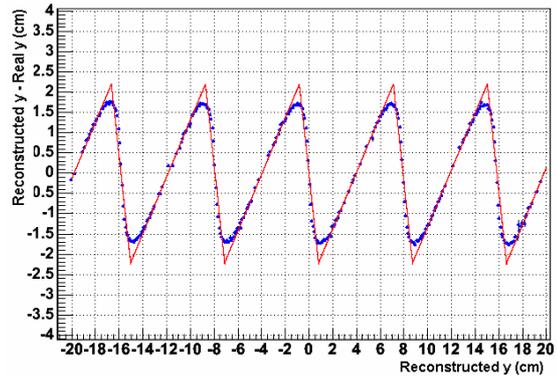

**Figure 7.** Reconstructed $y_c$ position vs the real photon incident $y$ position.

**Figure 8.** The $y_c$-$y$ variable as a function of the reconstructed photon position $y_c$.

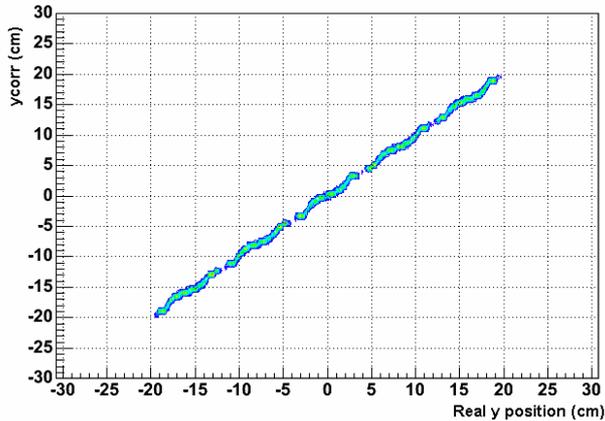

**Figure 9.** Photon $y_{corr}$ position after the correction procedure as a function of the real incident $y$ position.

4.5 "Hole" and Thyroid Phantoms:

After validating the simulation through comparison with experimental measurements, imaging tests were performed using two different phantoms: a phantom with variable size holes and a thyroid phantom.

A polymethylmethacrylate (PMMA) phantom consisting of seven holes with 3 cm depth and different diameters (2, 4, 6, 8, 10, 14 and 19 mm) was used (figure 10). The holes were filled with a $^{99m}$Tc solution of $3.3 \times 10^7$ Bq activity. The activity was distributed according to the volume of each hole. The phantom was placed at the collimator surface centered in the FOV and the same configuration was used in the simulation. About 12 million photons were generated according to the volumetric isotropic distribution for each hole. The image of phantom obtained experimentally was then compared to the image resulting from the corresponding simulation.

A thyroid phantom featuring structures typical of abnormal thyroids, hot and cold nodules regions of different activities and two areas with reduced activities, was used for this study (figure 11). The phantom was filled with a uniform solution of $^{99m}$Tc with an activity of $3.7 \times 10^7$ Bq and placed close to collimator surface centered on the FOV. For simulations an ellipsoid model of the thyroid based on the real test phantom was used as a geometric approximation of the real object.

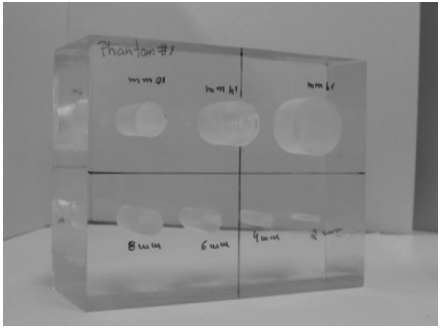

**Figure 10.** Photo of PMMA "hole" phantom. The holes have the following diameters: 2, 4, 6, 8, 10, 14 and 19 mm.

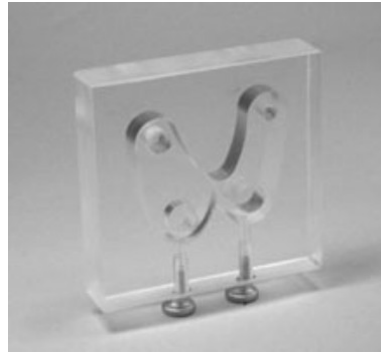

**Figure 11.** Photo of the thyroid phantom.

About 138 million photons were generated in the simulation with isotropic distribution within the volumes corresponding to the hot regions. The images of simulated phantom and experimental phantom were then compared. The imaging results obtained for the "hole" and the thyroid phantoms are presented in figure. 12 and figure 13, respectively. Both simulated and experimental images are shown side to side for easier comparison. The best results were obtained for the "hole" phantom, probably due to the oversimplified model used for the simulation of the thyroid phantom.

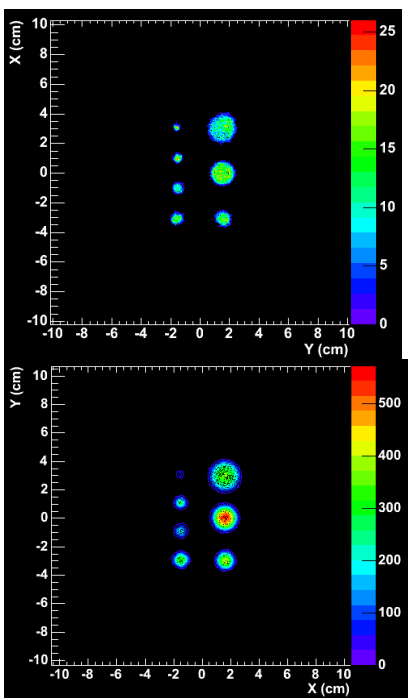

**Figure 12.** Simulated (upper) and experimental (lower) images of a phantom consisting of seven holes filled with $^{99m}$Tc solutions of different activity.

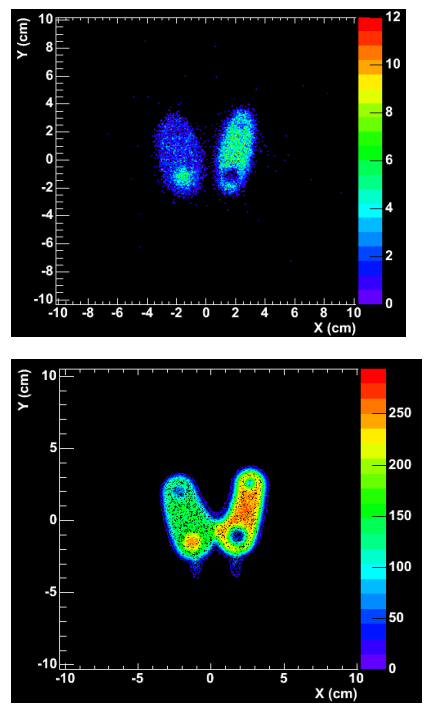

**Figure 13.** Simulated (upper) and experimental (lower) images of the numerical model and the experimental thyroid phantoms, respectively.

## 5. Conclusions

We have implemented a Geant4 based simulation of a Siemens E.Cam Dual Head gamma camera and performed its experimental characterization in terms of energy resolution, sensitivity, spatial resolution, linearity and imaging of phantoms using $^{99m}$Tc. The comparison between simulation results and experimental data allowed for the validation of the software codes developed for this work and provided a better understanding of the operation of the camera at detector level. These codes may now be used for modeling and developing nuclear medicine detector technology and thus contribute to improve the state-of-the-art in the field.

## Acknowledgments

The authors wish to thank the Unidade de Intervenção Cardiovascular, located at Hospital Particular do Algarve, Siemens Medical Solutions, Mechanical Workshop of Physics Department and M.D. Marília Pedrosa for the valuable collaboration. Usefull discussion with P. Sousa is gratefully acknowledged. This work was supported by the Portuguese Foundation for Science and Technology (FCT) and FEDER found under grants POCTI/FNU/47678/2002, POCTI/FP/63448/2005, BPD/5726/2001 and SFRH/BPD/11547/2002 and by the University of Algarve, Portugal.